\documentclass[letterpaper,11pt]{report}

\title
{
	Cryptolysis v.0.0.1 - A Framework for Automated
	Cryptanalysis of Classical Ciphers
}

{\author
	{\bf
		{CIISE Security Investigation Initiative}\\\hline\\
		Represented by:\\\\
		Serguei A. Mokhov\\
		Marc-Andr\'e Laverdi\`ere\\
		Nader Hatami\\
		Ali Benssam\\\\
		\texttt{\{mokhov,ma\_laver,nade\_hat,al\_ben\}@ciise.concordia.ca}
		\\\\\\
		Montr\'eal, Qu\'ebec, Canada\\\\\\
	}
}

\date{Fri  7 Jan 2011 15:21:19 EST
}

\usepackage{graphicx}
\usepackage{latexsym}
\usepackage{makeidx}
\usepackage{url}

\makeindex

\newcommand{\xf}[1]{Figure~\ref{#1}}

\newcommand{\java}{{Java\index{Java}}}

\newcommand{\todo}[0]
{
	{\Large \[TODO\]}
}

\newcommand{\api}[1]{\texttt{#1}\index{API!#1}}

\newcommand{\marf}[0]{MARF\index{Tools!MARF}\index{Libraries!MARF}}

\newcommand{\lucidL}[1]{{$\mathit{Lucid}$}($L$) }

		{}

\def\myvert{\raise 2.27pt \hbox{\vrule depth 0pt height 8pt width 0.2mm}}
\def\myarrow{\hspace*{0.43mm}%
             \raise 2.29pt\hbox{\vrule depth 0pt height 8pt width 0.16mm}%
             \hspace*{-0.32mm}%
             $\longrightarrow$
             \ %
             }

\begin{document}

	\begin{titlepage}
		\maketitle
	\end{titlepage}

	\pagenumbering{roman}
\tableofcontents
\clearpage
\pagenumbering{arabic}

\listoffigures
\listoftables

	\chapter{Introduction}
\index{Introduction}

$Revision: 1.2 $

\section{What is Cryptolysis?}

Cryptolysis is a framework that includes a collection of automated attacks
on the classical ciphers based on the article \cite{optheuristicscrypto}.

\section{Tools}

\subsection{{\java}}

We have chosen to implement our project using the Java programming language. This
choice is justified by the binary portability of the Java applications as well as
facilitating memory management tasks and other issues, so we can concentrate more on
the algorithms instead. Java also provides
us with built-in types and data-structures to manage
collections (build, sort, store/retrieve) efficiently \cite{javanuttshell}.

\subsection{{\marf}}

Portions of Cryptolysis reply on {\marf}, described in many published works,
we just cite one instance here for the follow up information \cite{marf}.



	\chapter{Design and Architecture}
\index{Architecture}

$Revision: 1.2 $

Before we begin, you should understand the basic
system architecture. Understanding how the
parts interact will make the follow up sections
somewhat clearer. This document presents architecture
of the Cryptolysis system, including the layout of the physical
directory structure, and Java packages.

\xf{fig:packages-design} lists the Java packages, \xf{fig:ciphers-design}
presents the Ciphers Framework; its purpose is to have some quick in-house
tools for creating testing ciphertexts. The most common API in here
is a series of \api{encrypt()} and \api{decrypt()} calls.
Then, \xf{fig:analyzers-design} presents the core of this work, namely
modules to perform the described attacks on ciphertext. They all
implement the \api{analyze()} method. In \xf{fig:app-design}, is the
way the main application uses (i.e. instantiates) the concrete modules
based on the set of options supplied.

\begin{figure}
	\centering
	\includegraphics[width=\textwidth]{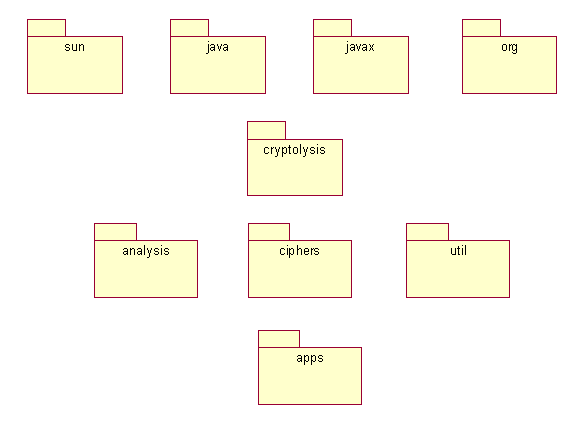}
	\caption{Cyrptolysis Packages}
	\label{fig:packages-design}
\end{figure}

\begin{figure}
	\centering
	\includegraphics[width=\textwidth]{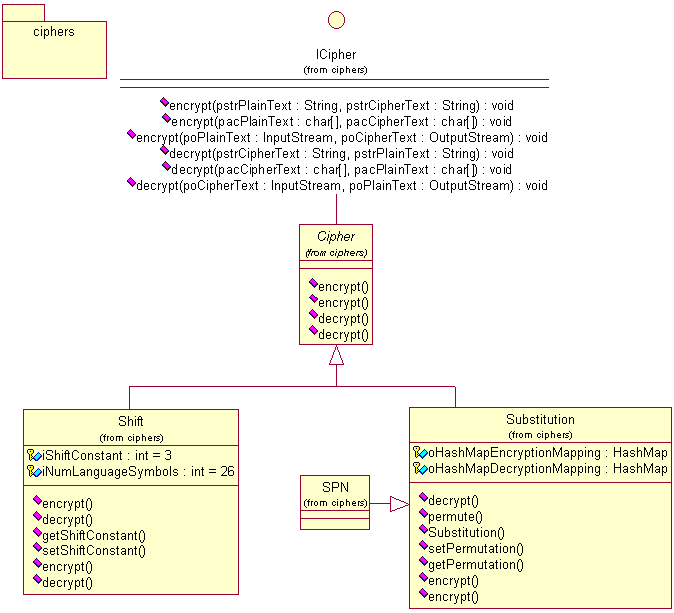}
	\caption{Ciphers Framework}
	\label{fig:ciphers-design}
\end{figure}

\begin{figure}
	\centering
	\includegraphics[width=\textwidth]{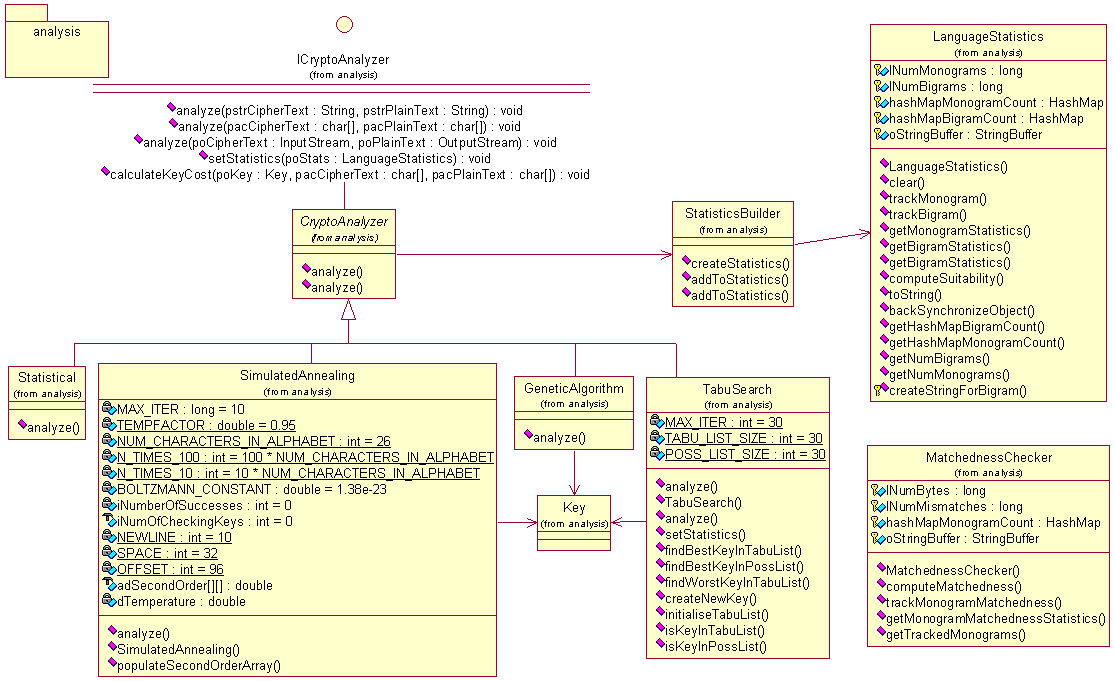}
	\caption{Crypto Analyzers Framework}
	\label{fig:analyzers-design}
\end{figure}

\begin{figure}
	\centering
	\includegraphics[width=\textwidth]{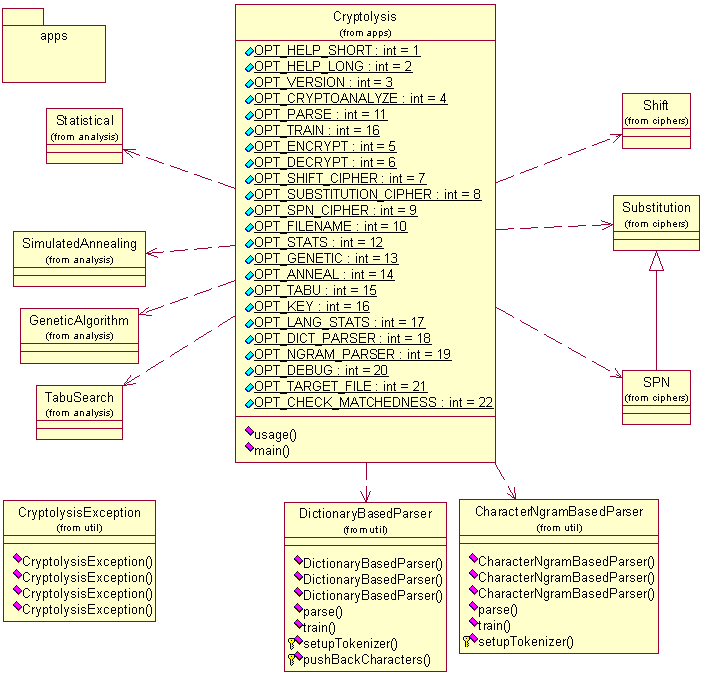}
	\caption{Use of Modules By the Cryptolysis Application}
	\label{fig:app-design}
\end{figure}

{\todo}


	\chapter{Methodology}

$Revision: 1.4 $

\section{Cryptanalysis Heuristics}

Most of this work is based on \cite{optheuristicscrypto}.

%
%

\section{Word Boundary Detection}

Since the deciphered text comes in as a stream of characters
with no spaces or punctuation as a bonus it is a good idea
to implement automatic placement of spaces between words
in the text for ease of reading. This problem in speech recognition and natural
language processing (NLP) is often referred to as {\em word boundary detection}.
The two main methods approaching this problem of converting a stream of characters
to a stream of words are a dictionary approach
and a statistical approach of n-gram language models. The models can be
further refined with sentence boundary detection via statistical NLP parsing.
The below are the details of the methodology for implementation of these two approaches in Cryptolysis,
their current paramters, advantages, limitations, and the way these
limiations can be overcome.

All methods have some training to do prior use on the source language
corpora\footnote{A {\em corpora} is a collection of natural language
documents used for any sort of natural language processing technique.}.
The trained data has to be serialized somehow and be usable
at later runs of the application.

\subsection{Dictionary Approach}

The dictionary approach is the most common one.
One has to compile a dictionary of words of English
from some training corpora. Next, when doing the word
boundary detection, start with the first character
of the stream and try to match a longest possible
word that begins with this character and subsequent
characters match; if so insert a space.

\paragraph*{The Requirements}

\begin{itemize}
\item
If not importing the dictionary from some external
sources, the training corpora must be large to include
more words.

\item
Recovery technique should be in place if a matching word
is not found in the dictionary, e.g. by proceeding further
until a word is found and keep the previous sequence of
characters as if it was a word.
\end{itemize}

\paragraph*{Problems and Limitations}

\begin{itemize}
\item
Word is not found; especially this is pertinent to
to proper names of people, places, etc. which all cannot
be easily found in the training copora in comprehensive
manner. Likewise, some domain specific terms and names, equations,
etc. may cause this problem.

\item
Composite words is another issue. By matching the longest
possible matching sequence may miss a space or cases
and won't be able to distinguish whether there was
a space or not. For example, these two cases are legal:
``{\em therefore}'' and ``{\em there for eternity}''
in example setences ``{\em Therefore, the theorem holds.}''
and ``{\em This stone was lying there for eternity.}''. Both
instances will be picked up as ``{\em therefore}'' resuling
in a non-dictionary word ``{\em ternity}'' for the second
example. Similar problem applies to ``{\em thereto}'',
``{\em thereafter}'', ``{\em thereby}'', and other composite words.

\item
The storage space required to contain the dictionary is large
and keeps growing if you add more words at training. Therefore,
the look up speed reduces.
\end{itemize}

\subsection{Character N-gram Language Model}

Unlike the dictionary approach, the character $n$-gram model
looks at sequences of $n$ characters when looking up spaces; thus,
it operates on the source character stream directly and
produces the word stream from probabilistic table look ups.
This is a statistical approach. Here we chose a 3-gram model
for space detection. During training for each word we count
how often the following occurs:

\begin{itemize}
\item
2 last characters of the preceeding word and a space
\item
last character of the preceeding word, space, and the first character of the second word 
\item
space and the first two characters of the second word
\end{itemize}

The frequencies later on stored in the probabilistic table
and the table is serialized. Of course, text boundaries are need
to be accounted for as there is no preceeding word before first at the beginning
and the second word after last.

These 2-character sequences (barring space) are then looked up
in the text to parse after the deciphering and spaces are put
according to the probabilites found during training.

The storage space requrires for this method are a lot less than
that of the dictionary approach and scanning is faster. However,
it may put spaces in the undesired places due to irregularities
in natural language.

\paragraph*{Problems}

\begin{itemize}
\item
One-character words
\item
Creating non-existent words
\end{itemize}

\subsection{Statistical Parsing}

Statistical NLP parsing \cite{jurafsky,cyk,marf} can help disambiguate with the composite words
and even find the sentence boundaries. This can be used as refinement tool
for either for the methods presented above. By trying to parse a longest parseable
span of words with a valid parse would give the setence boundary. If no parse
at all found, it either means the word boundary for some words was not done
properly or the words are not in the dicitonary/grammar or the source
text has not properly English-formed sentences.

{\todo}


	\chapter{Applications}
\label{chapt:apps}
\index{Applications}

$Revision: 1.2 $

This chapter is to describe the application that employs 
the Cryptolysis framework. Its source code revision 
is quoted in the Appendix and is subject to maintenance
on SourceForge as resources permit.

{\todo}


	\chapter{Results}
\label{sect:results}
\index{Results}

$Revision: 1.2 $

The results are found in the aux folder in the arXiv submissions
and have not been integrated yet as nice tables and graphs into
the document. Please consult the files in that directory.



	\chapter{Conclusions}

$Revision: 1.2 $

We have built a simple framework for verification of classical cipher
attacks, that is expandable and a platform for comparative experiments.
We obtained some encouraging results by guessing out the keys during
the implemented attack scenarios. We plan on expanding to include
other algorithms and frameworks in the testing environment and 
improve integration with other frameworks.


\section{Acknowledgments}

Dr. Amr M. Youssef and Faculty of Engineering and Computer Science, Concordia University,
Montreal, Canada.



	\addcontentsline{toc}{chapter}{Bibliography}

\bibliography{report}
\bibliographystyle{alpha}


	\appendix

\chapter{\api{Cryptolysis} Application Source Code}

\vspace{15pt}
\hrule
{\scriptsize \begin{verbatim}
package cryptolysis.apps;

import java.io.FileInputStream;
import java.io.InputStream;
import java.io.FileOutputStream;
import java.io.OutputStream;

import marf.MARF;
import marf.Storage.IStorageManager;
import marf.util.Debug;
import marf.util.OptionProcessor;
import cryptolysis.analysis.GeneticAlgorithm;
import cryptolysis.analysis.ICryptoAnalyzer;
import cryptolysis.analysis.LanguageStatistics;
import cryptolysis.analysis.MatchednessChecker;
import cryptolysis.analysis.SimulatedAnnealing;
import cryptolysis.analysis.Statistical;
import cryptolysis.analysis.StatisticsBuilder;
import cryptolysis.analysis.TabuSearch;
import cryptolysis.ciphers.ICipher;
import cryptolysis.ciphers.SPN;
import cryptolysis.ciphers.Shift;
import cryptolysis.ciphers.Substitution;
import cryptolysis.util.CharacterNgramBasedParser;
import cryptolysis.util.CryptoParser;
import cryptolysis.util.CryptolysisException;
import cryptolysis.util.DictionaryBasedParser;


/**
 * <p>Main Cryptolysis Application.</p>
 * 
 * TODO: document
 * 
 * $Id: Cryptolysis.java,v 1.19 2005/11/10 10:25:22 mokhov Exp $
 * 
 * @author Serguei Mokhov
 * @author Marc-Andre Laverdiere
 */
public class Cryptolysis
{
    // A bunch of options
    // TODO: document
    
    public static final int OPT_HELP_SHORT = 1;
    public static final int OPT_HELP_LONG = 2;
    public static final int OPT_VERSION = 3;

    public static final int OPT_CRYPTOANALYZE = 4;
    public static final int OPT_PARSE = 11;
    public static final int OPT_TRAIN = 16;

    public static final int OPT_ENCRYPT = 5;
    public static final int OPT_DECRYPT = 6;
    
    public static final int OPT_SHIFT_CIPHER = 7;
    public static final int OPT_SUBSTITUTION_CIPHER = 8;
    public static final int OPT_SPN_CIPHER = 9;
    public static final int OPT_KEY = 16;

    public static final int OPT_FILENAME = 10;

    public static final int OPT_STATS = 12;
    public static final int OPT_GENETIC = 13;
    public static final int OPT_ANNEAL = 14;
    public static final int OPT_TABU = 15;

    public static final int OPT_LANG_STATS = 17;
    public static final int OPT_DICT_PARSER = 18;
    public static final int OPT_NGRAM_PARSER = 19;

    public static final int OPT_DEBUG = 20;
    
    public static final int OPT_TARGET_FILE = 21;
    public static final int OPT_CHECK_MATCHEDNESS = 22;
    
    /**
     * @param argv command-line parameters
     */
    public static final void main(String[] argv)
    {
        try
        {
            //Debug.enableDebug(true);

            OptionProcessor oGetOpt = new OptionProcessor();
    
            oGetOpt.addValidOption(OPT_HELP_SHORT, "-h");
            oGetOpt.addValidOption(OPT_HELP_LONG, "--help");
            oGetOpt.addValidOption(OPT_VERSION, "--version");
            oGetOpt.addValidOption(OPT_DEBUG, "--debug");
            
            oGetOpt.addValidOption(OPT_CRYPTOANALYZE, "--analyze");
            oGetOpt.addValidOption(OPT_ENCRYPT, "--encrypt");
            oGetOpt.addValidOption(OPT_DECRYPT, "--decrypt");
            oGetOpt.addValidOption(OPT_PARSE, "--parse");
            oGetOpt.addValidOption(OPT_TRAIN, "--train");
            
            oGetOpt.addValidOption(OPT_SHIFT_CIPHER, "-shift");
            oGetOpt.addValidOption(OPT_SUBSTITUTION_CIPHER, "-subst");
            oGetOpt.addValidOption(OPT_SPN_CIPHER, "-spn");
    
            oGetOpt.addValidOption(OPT_STATS, "-stats");
            oGetOpt.addValidOption(OPT_ANNEAL, "-anneal");
            oGetOpt.addValidOption(OPT_GENETIC, "-genetic");
            oGetOpt.addValidOption(OPT_TABU, "-tabu");

            oGetOpt.addValidOption(OPT_LANG_STATS, "-lang");
            oGetOpt.addValidOption(OPT_DICT_PARSER, "-dict");
            oGetOpt.addValidOption(OPT_NGRAM_PARSER, "-ngram");
            
            oGetOpt.addValidOption(OPT_CHECK_MATCHEDNESS,"--checkmatch");
            
    
            int iValidOptions = oGetOpt.parse(argv);
            
            if(oGetOpt.isActiveOption(OPT_DEBUG))
            {
                Debug.enableDebug();
            }

            if
            (
                oGetOpt.isActiveOption(OPT_HELP_SHORT)
                || oGetOpt.isActiveOption(OPT_HELP_LONG)
            )
            {
                usage();
                System.exit(0);
            }
            
            if(oGetOpt.isActiveOption(OPT_VERSION))
            {
                System.out.println
                (
                    "Cryptolysis $Revision: 1.19 $\n"
                    + "Using MARF v." + MARF.getVersion()
                );
                
                System.exit(0);
            }
            
            switch(oGetOpt.getInvalidOptions().size())
            {
                case 0:
                    break;
                
                // Shall be the filename
                case 1:
                {
                    oGetOpt.addActiveOption(OPT_FILENAME, oGetOpt.getInvalidOptions().firstElement().toString());
                    oGetOpt.getInvalidOptions().clear();
                    break;
                }

                // Shall be the filename and the key    
                case 2:
                {
                    if(oGetOpt.isActiveOption(OPT_CHECK_MATCHEDNESS))
                    {
                        oGetOpt.addActiveOption(OPT_FILENAME, oGetOpt.getInvalidOptions().firstElement().toString());
                        oGetOpt.addActiveOption(OPT_TARGET_FILE, oGetOpt.getInvalidOptions().elementAt(1).toString());
                        oGetOpt.getInvalidOptions().clear();
                        
                    } 
                    else
                    {
                        oGetOpt.addActiveOption(OPT_FILENAME, oGetOpt.getInvalidOptions().firstElement().toString());
                        oGetOpt.addActiveOption(OPT_KEY, oGetOpt.getInvalidOptions().elementAt(1).toString());
                        oGetOpt.getInvalidOptions().clear();
                    }
                    break;
                }

                case 3:
                    oGetOpt.addActiveOption(OPT_FILENAME, oGetOpt.getInvalidOptions().firstElement().toString());
                    oGetOpt.addActiveOption(OPT_KEY, oGetOpt.getInvalidOptions().elementAt(1).toString());
                    oGetOpt.addActiveOption(OPT_TARGET_FILE, oGetOpt.getInvalidOptions().elementAt(2).toString());

                    oGetOpt.getInvalidOptions().clear();
                    
                    break;
                    
                    

                
                default:
                    throw new CryptolysisException
                    (
                        "Invalid options found: " + oGetOpt.getInvalidOptions()
                    );
            }
            
            InputStream oInputText = null;
            InputStream oInputTextCompare = null;
            OutputStream oOutputText = null;
            
            // Get input from either a file name if specified, or STDIN
            if(oGetOpt.isActiveOption(OPT_FILENAME))
            {
                oInputText = new FileInputStream(oGetOpt.getOption(OPT_FILENAME));
            }
            else
            {
                oInputText = System.in;
            }
            
            if(!oGetOpt.isActiveOption(OPT_CHECK_MATCHEDNESS) &&
                oGetOpt.isActiveOption(OPT_TARGET_FILE))
            {
                oOutputText = new FileOutputStream(oGetOpt.getOption(OPT_TARGET_FILE));
            }
            else
            {
                oOutputText = System.out;
            }
            if (oGetOpt.isActiveOption(OPT_CHECK_MATCHEDNESS) && 
                oGetOpt.isActiveOption(OPT_TARGET_FILE)    )
            {
                oInputTextCompare = new FileInputStream(oGetOpt.getOption(OPT_TARGET_FILE));
            }
                

            // Assume defaults
            if(iValidOptions == 0)
            {
                // Run statistical cryptanalysis assuming STDIN
                oGetOpt.addActiveOption(OPT_CRYPTOANALYZE, "--analyze");
                oGetOpt.addActiveOption(OPT_STATS, "-stats");
            }

            Debug.debug("Final set of options: " + oGetOpt);
    
            if(oGetOpt.isActiveOption(OPT_CRYPTOANALYZE))
            {
                ICryptoAnalyzer oCryptanalysis = null;

                // Pick a heuristic strategy
                if(oGetOpt.isActiveOption(OPT_ANNEAL))
                {
                    oCryptanalysis = new SimulatedAnnealing();
                }
                else if(oGetOpt.isActiveOption(OPT_TABU))
                {
                    oCryptanalysis = new TabuSearch();
                }
                else if(oGetOpt.isActiveOption(OPT_GENETIC))
                {
                    oCryptanalysis = new GeneticAlgorithm();
                }
                else
                {
                    oCryptanalysis = new Statistical();
                }

                // Parse
                if(oGetOpt.isActiveOption(OPT_PARSE))
                {
                    Debug.debug("Parsing text...");

                    //String strOutfile = oCryptanalysis.getClass().getName() + ".parsed.txt";
                    //FileOutputStream oFOS = new FileOutputStream(strOutfile);
                    //oCryptanalysis.analyze(oInputText, oFOS);
                    //oFOS.close();
                    
                    //FileInputStream oFIS = new FileInputStream(strOutfile);

                    FileInputStream oFIS = new FileInputStream(oGetOpt.getOption(OPT_FILENAME));
                    CryptoParser oParser = new DictionaryBasedParser(oFIS);
                    oParser.parse();
                }
                else
                {
                    Debug.debug("Analyzing text...");
                    
                    // Re-load stored trained stats.
                    LanguageStatistics oStats = new LanguageStatistics();

                    oStats.setDumpMode(IStorageManager.DUMP_GZIP_BINARY);
                    oStats.setFilename(oStats.getClass().getName() + "." + oStats.getDefaultExtension());
                    oStats.restore();

                    //Debug.debug(oStats);
                    oCryptanalysis.setStatistics(oStats);
                    oCryptanalysis.analyze(oInputText, System.out);
                }
            }


            /*
             * Encryption
             */
            else if(oGetOpt.isActiveOption(OPT_ENCRYPT))
            {
                ICipher oCipher = null;

                if(oGetOpt.isActiveOption(OPT_SPN_CIPHER))
                {
                    oCipher = new SPN();
                }
                else if(oGetOpt.isActiveOption(OPT_SUBSTITUTION_CIPHER))
                {
                    oCipher = new Substitution();
                    
                    //prepare key
                    String strPermutationKey = oGetOpt.getOption(OPT_KEY);
                    char[] acKey = strPermutationKey.toCharArray();
                    char[] acOriginal = new char[acKey.length];
                    
                    //Assumes English characters only for now
                    int iBase = 'A';

                    for(int i = 0; i < acKey.length; i++)
                    {
                        acOriginal[i] = (char)(iBase + i);
                    }
                    
                    ((Substitution) oCipher).setPermutation(acOriginal, acKey);
                }
                else
                {
                    oCipher = new Shift();
                }

                //System.out.println("Encrypting text...");
                oCipher.encrypt(oInputText, oOutputText);
                //System.out.println("Encryption done.");
            }
            else if(oGetOpt.isActiveOption(OPT_DECRYPT))
            {
                ICipher oCipher = null;

                if(oGetOpt.isActiveOption(OPT_SPN_CIPHER))
                {
                    oCipher = new SPN();
                }
                else if(oGetOpt.isActiveOption(OPT_SUBSTITUTION_CIPHER))
                {
                    oCipher = new Substitution();
                }
                else
                {
                    oCipher = new Shift();
                }
                
                oCipher.decrypt(oInputText, oOutputText);
            }


            /*
             * Training
             */
            else if(oGetOpt.isActiveOption(OPT_TRAIN))
            {
                Debug.debug("Training...");

                if(oGetOpt.isActiveOption(OPT_LANG_STATS))
                {
                    Debug.debug("Training language statistics.");

                    StatisticsBuilder oBuilder = new StatisticsBuilder();
                    LanguageStatistics oStats = oBuilder.createStatistics();

                    oBuilder.addToStatistics(oStats, oInputText);

                    // Serialize stats data to be reloaded
                    // in the future when running an attack.
                    // It is dumped gzip compressed with the name of the
                    // class as a filename with a .gzbin extension.
                    oStats.setDumpMode(IStorageManager.DUMP_GZIP_BINARY);
                    oStats.setFilename(oStats.getClass().getName() + "." + oStats.getDefaultExtension());
                    oStats.dump();
                    
                    // Prints out a text prepresentation of the language statistics object
                    System.out.println(oStats);
                }
                else if(oGetOpt.isActiveOption(OPT_DICT_PARSER))
                {
                    Debug.debug("Training dictionary-based parser.");
                    DictionaryBasedParser oParser = new DictionaryBasedParser(oInputText);
                    oParser.train();
                }
                else if(oGetOpt.isActiveOption(OPT_NGRAM_PARSER))
                {
                    Debug.debug("Training N-gram-based parser.");
                    CharacterNgramBasedParser oParser = new CharacterNgramBasedParser(oInputText);
                    oParser.train();
                }
                else
                {
                    throw new Exception("No valid training module type found.");
                }

                Debug.debug("Training done.");
            }
            
            /*
             * Check matchedness
             */
            else if(oGetOpt.isActiveOption(OPT_CHECK_MATCHEDNESS)){
                MatchednessChecker oChecker = new MatchednessChecker();
                double result = oChecker.computeMatchedness(oInputText,oInputTextCompare);
                System.out.println( oGetOpt.getOption(OPT_FILENAME) + " vs " + 
                                    oGetOpt.getOption(OPT_TARGET_FILE) + "\n\tmatching at : "+
                                    result*100 + " %");
                
                System.out.println("Statistics for each monogram: ");
                char[] monograms = oChecker.getTrackedMonograms();
                for (int i = 0; i < monograms.length; i++)
                {
                    System.out.println("\t" + monograms[i] +": " + 
                            oChecker.getMonogramMatchednessStatistics(monograms[i]) * 100 + "%" );
                    
                }
                
            }
        }
        catch(Exception e)
        {
            System.err.println(e.getMessage());
            e.printStackTrace(System.err);
            System.exit(1);
        }
    }
    
    public static final void usage()
    {
        System.out.println
        (
            "Cryptolysis $Revision: 1.19 $\n"
            + "A Crytanalysis Framework for Classical Ciphers\n"
            + "Author: CIISE Security Investigation Iniative\n\n"
            
            + "Usage:\n\n"
            + "   java Cryptolysis --help | -h\n"
            + "       displays usage information\n\n"
            
            + "   java Cryptolysis --version\n"
            + "       displays version information\n\n"
            
            + "   java Cryptolysis [ OPTIONS ] [ --debug ]\n"
            + "       does cryptoanalysis and cryptography-related tasks\n\n"
            
            + "Where options are one of the following:\n\n"
            + "   --analyze [ --parse ] [ ATTACK ] [ FILENAME ]\n"
            + "   --encrypt [ CIPHER ] [ FILENAME KEY TARGETFILE ]\n"
            + "   --decrypt [ CIPHER ] [ FILENAME KEY TARGETFILE ]\n"
            + "   --train STATSMODULE [ FILENAME ]\n"
            + "   --checkmatch FILENAME FILENAME \n\n"
            
            + "Where ATTACK is one or more of the following:\n\n"
            + "   -stats    Statistical N-gram Model Attack\n"
            + "   -genetic  Genetic Algorithm Attack\n"
            + "   -anneal   Simulated Annealing Attack\n"
            + "   -tabu     Tabu Search Attack\n\n"

            + "Where CIPHER is one or more of the following:\n\n"
            + "   -shift    Shift Cipher\n"
            + "   -subst    Substitution Cipher\n"
            + "   -spn      SPN Cipher\n\n"

            + "Where STATSMODULE is one of the following:\n\n"
            + "   -dict     Dictionary-based\n"
            + "   -ngram    Character N-gram based\n"
            + "   -lang     Language N-gram statistics\n"
        );
    }
}

// EOF
\end{verbatim}
}
\hrule
\vspace{15pt}




	\printindex
\end{document}